\documentclass[%
 reprint,
superscriptaddress,
groupedaddress,
 amsmath,amssymb,
 aps,
floatfix,
]{revtex4-2}

\usepackage{graphicx}
\usepackage{dcolumn}
\usepackage{bm}
\usepackage[utf8x]{inputenc}

\begin{document}

\preprint{APS/123-QED}

\title{Transverse momentum spectra and suppression of charged hadrons in deformed Xe-Xe collisions at $\sqrt{s_{NN}}$ = 5.44 TeV using HYDJET++ model}

\author{Saraswati Pandey}
 \affiliation{Department of Physics, Institute of Science, Banaras Hindu University (BHU), Varanasi, 221005, INDIA.}
 \email{saraswati.pandey13@bhu.ac.in \\
 bksingh@bhu.ac.in}

\author{B. K. Singh\footnote{bksingh@bhu.ac.in}}%
 \affiliation{Department of Physics, Institute of Science, Banaras Hindu University (BHU), Varanasi, 221005, INDIA.}

\date{\today}

\begin{abstract}
\noindent
In this study, we systematically investigate deformed Xe-Xe collisions at 5.44 TeV center of mass energy. We exploit Monte Carlo HYDJET++ model to compute transverse momentum ($p_{T}$) distribution, nuclear modification factor $R_{AA}$ and relative suppression in terms of $R_{CP}$ as a function of transverse momentum and centrality of collision of charged hadrons in body-body and tip-tip geometrical configurations, respectively. We have compared HYDJET++ model results to those from ALICE experimental data and AMPT model (String-Melting version) results. Minimum bias Xe-Xe collisions show a suitable match with ALICE experimental data at midrapidity. Average transverse momentum ($\langle p_{T} \rangle$) show strong centrality dependence. The contribution of hard parton scatterings to the total charged hadron yield is discussed. Both nuclear modification factor $R_{AA}$ and relative suppression $R_{CP}$ show clear dependence on centrality as well as transverse momentum. The charged hadron observables show clear dependence on the geometrical configuration of the collisions. HYDJET++ model justifies experimental data more closely than AMPT model.   
\end{abstract}

\date{\today}
\maketitle 

\section{Introduction}
\label{intro}
One of the primary mission of ultra-relativistic nuclear collisions as performed at the Relativistic Heavy-Ion Collider (RHIC) and the Large Hadron Collider (LHC), is to produce and study, a novel and peculiar state of matter called as Quark-Gluon plasma (QGP) and its properties \cite{doi:10.1146/annurev-nucl-101917-020852}. High transverse momentum quarks and gluons (jets), created from early stage hard scatterings are very useful probes of such highly excited nuclear matter \cite{Mulligan:2020bqi}. These hard partonic jets lose energy as they interact with the medium ingredients while propagating through the QGP medium before fragmenting into hadrons. So, the associated observables are modified. Transverse momentum $p_{T}$ spectra of charged hadrons is a suitable observable for studying these hard partonic jets \cite{QIN2014165}.

Between low to intermediate $p_{T}$ range, collective expansion of the system rules the production of charged hadrons. This is evident from single-particle $p_{T}$ spectra and multi-particle correlations \cite{PhysRevC.93.034913, doi:10.1146/annurev-nucl-102212-170540}. At high $p_{T}$, particles emanate from parton fragmentation and are susceptible to the energy loss partons suffer while propagating in the medium \cite{2019166}. According to a simplified model, the energy loss depends on the number of scattering centers, being approximately proportional to the energy density, and on the parton path length propagated in the medium \cite{Bjorken:1982tu}. The relationship is linear for elastic collisions and quadratic for medium induced gluon radiation \cite{dEnterria:2009xfs, PhysRevC.97.014910}. One of the major signals of QGP formation is the strong suppression of these high-$p_{T}$ particles (called as jet quenching) observed in heavy ion collisions at RHIC and LHC. Energy loss is dominated by the radiative mechanism by way of induced gluon emission for fast partons \cite{Zakharov:2021uza}. At LHC, the suppression of particle production in Pb-Pb collisions is observed to be stronger by a factor of 7-8 for $p_{T}$ around 6-7 GeV/c, linearly decreasing at high $p_{T}$ and ample suppression above 100 GeV/c. Charged particle yields at high $p_{T}$ are weakly modified in smaller collision systems, thereby excluding the effects related to initial-state situations of the nucleus caused by high-$p_{T}$ suppression \cite{ALICE:2018vuu, CMS:2016xef}. Altogether, these observations imply strong $p_{T}$-dependent energy loss in QGP environment of heavy-ion collisions.

LHC in 2017, collided Xenon by an eight-hour run at 5.44 TeV center-of-mass energy. Initially, it only provided proton-proton (pp), p-Pb, and Pb-Pb collisions. Xenon nucleus has a mass number lying mid-between p and Pb$^{208}$. Therefore, colliding Xe$^{129}$ nuclei will bridge the multiplicity gap between larger Pb-ion systems and the smaller, p + p and p + Pb systems. In contrary to Pb nucleus (which is spherical in shape), Xe-nucleus is somewhat prolate in shape \cite{MOLLER20161}. The deformation in xenon allows us to probe a different initial condition. Therefore, Xe-Xe collisions will provide a unique opportunity to explore QGP and its properties using intermediate-size collision system at LHC energies. Assuming energy loss of parton being linearly or quadratically related to only path length through QGP indicate an average reduction in energy loss of 17\% or 31\% respectively, in head-on Xe-Xe collisions compared to Pb-Pb collisions \cite{2019166}.  Also, deformed shape brings multiple geometrical configurations in accordance to the way nuclei collide with each other such as body-body, body-tip, tip-tip, etc. Here we will focus on two geometrical configurations: body-body and tip-tip configurations, depending upon the angle, the colliding nuclei make with the reaction plane \cite{PhysRevC.85.034905}. The overlapping region in Xe-Xe collisions will surely not be circular, and therefore, the observables may show a change when measured in comparison to non-deformed collisions \cite{PhysRevC.103.014903}. Also, in deformed collisions, charged particle multiplicity density in the transverse phase space is expected to be higher than the spherical nucleus collisions \cite{PhysRevC.73.034911, TRIPATHY201881}. 

This paper presents the study of transverse momentum ($p_{T}$) spectra of charged hadrons with and without jet contribution in Xe-Xe collisions at 5.44 TeV center-of-mass energy under HYDJET++ model framework. The results produced have been analyzed in body-body and tip-tip geometrical configurations. The observables have been compared with the ALICE experimental data at LHC in the kinematic range 0.15$<p_{T}<$30 GeV/c and $|\eta|<$0.8 for seven classes of collision centrality. We have calculated nuclear modification factor $R_{AA}$ and suppression in terms of $R_{CP}$ for each centrality of collision. The study is organized as follows: In Sec. \ref{model}, we concisely discuss the formulation of HYDJET++ model and the deformation incorporated in the structure of the model. In Sec. \ref{j and jq}, we debate about Jet and Jet Quenching and its measurement in heavy-ion collisions. In Sec. \ref{results}, we present the results and discussions for transverse momentum $p_{T}$  distribution, nuclear modication factor $R_{AA}$ and suppression of $p_{T}$-spectra. Lastly, we have summarized our outcomes in Sec. \ref{summary}.

\section{Model Formalism}
\vspace{-2ex}
\label{model}
HYDJET++ (hydrodynamics plus jets) is an event generator developed to study heavy-ion collisions at RHIC as well as LHC energies, performing simulation by superimposing soft (hydro-type) state and the hard state (resulting from multi-parton fragmentation) simultaneously. These soft and hard parts are treated independently in HYDJET++. It meticulously treats soft hadroproduction (collective flow phenomenon and the resonance decays) as well as hard parton production, also examining the known medium effects (jet quenching and nuclear shadowing). Incorporated physics and the corresponding simulation procedure can be found in the articles \cite{LOKHTIN2009779, Lokhtin:2009be, bravina2017dynamical}. The imbibed physics of the model concisely is as follows:

\begin{figure}[htbp]
\centering
\includegraphics[scale=0.25]{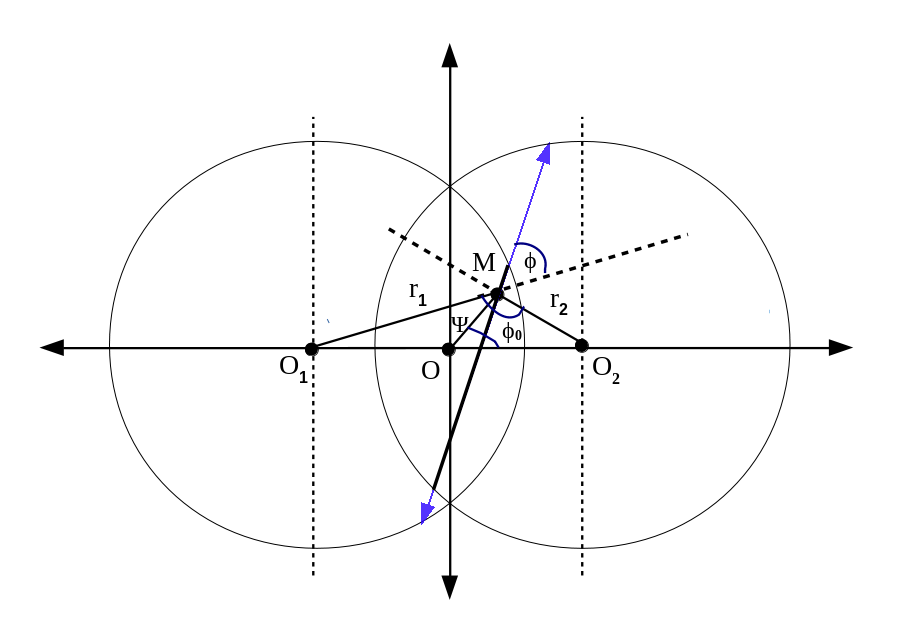} 
\caption{Jet production in a high energy symmetric A-A collision in impact parameter (b) plane. $O_{1}$ and $O_{2}$ are the nucleus centers, $OO_{2}$ = −$O_{1}$O = b/2. M($r \cos \psi, r \sin \psi$) is the jet (dijet) production vertex, r is the distance from the nuclear collision axis to M, $r_{1}$ , $r_{2}$ are the distances between the nucleus centers ($O_{1}$,$O_{2}$) and M; $\phi$ is the jet azimuthal angle, and $\phi_{0}$ is the azimuthal angle between the vectors $r_{1}$ and $r_{2}$ \cite{lokhtin2000nuclear}.}
\label{jet}
\end{figure}

\textbf{Hard State:} The hard multi-parton state of HYDJET++ event is similar to HYDJET event generator \cite{lokhtin2006model, Lokhtin:2007ga}. It is based on the PYQUEN (Pythia Quenched) partonic loss model \cite{lokhtin2006model}. PYTHIA is an event generator that simulates hard nucleon-nucleon (NN) collision considering only those events whose generated total transverse momentum $p_{T}$ is higher than $p_{T}^{min}$. PYQUEN generates binary nucleonic collision vertices according to the Glauber model at a certain impact parameter. In short, HYDJET++ uses PYQUEN which modifies a jet event generated by PYTHIA. $p_{T}^{min}$ is an input parameter in HYDJET++ which separates the soft part of the event from the hard part. It is the minimum transverse momentum transfer of hard parton-parton scatterings in GeV/c  \cite{LOKHTIN2009779}. Events for which generated $p_{T}<p_{T}^{min}$ are taken into the soft part of the model. Partons produced in (semi-)hard processes with a momentum transfer lower than $p_{T}^{min}$ are thermalized, therefore, their hadronization products are moved to the soft part of the event automatically. In the model framework, this $p_{T}^{min}$ parameter is used to calculate the mean number of jets produced in AA events at a given impact parameter b. Then event-by-event simulation followed by rescattering of parton path in the dense medium is carried out including the radiative and collision energy losses per rescattering \cite{PhysRevD.27.140, PhysRevD.44.R2625, lokhtin2000nuclear, PhysRevC.60.064902, PhysRevC.64.057902}. Finally, hadronization happens in the framework of the Lund string model for hard partons and in-medium emitted gluon  \cite{andersson2005lund}. Nuclear shadowing of the parton distribution function, an important cold nuclear matter effect is included using an impact parameter-dependent parametrization performed in the framework of the Glauber-Gribov theory \cite{TYWONIUK2007170, gribov1969glauber}. An important point to be noted is that the collisional energy loss due to scattering with low momentum transfer is not included in HYDJET++ because its contribution to the total collisional energy loss is very small compared to high momentum scattering. Medium, where parton rescattering occurs, is a boost-invariant longitudinally expanding quark-gluon fluid. Partons are produced on a hypersurface of equal proper times $\tau$. HYDJET++ uses Bjorken hydrodynamics, therefore, cannot be applied at larger rapidities where Landau hydrodynamics are more suitable for the proper description of medium expansion.

\begin{figure}[htbp]
\centering
\includegraphics[scale=0.15]{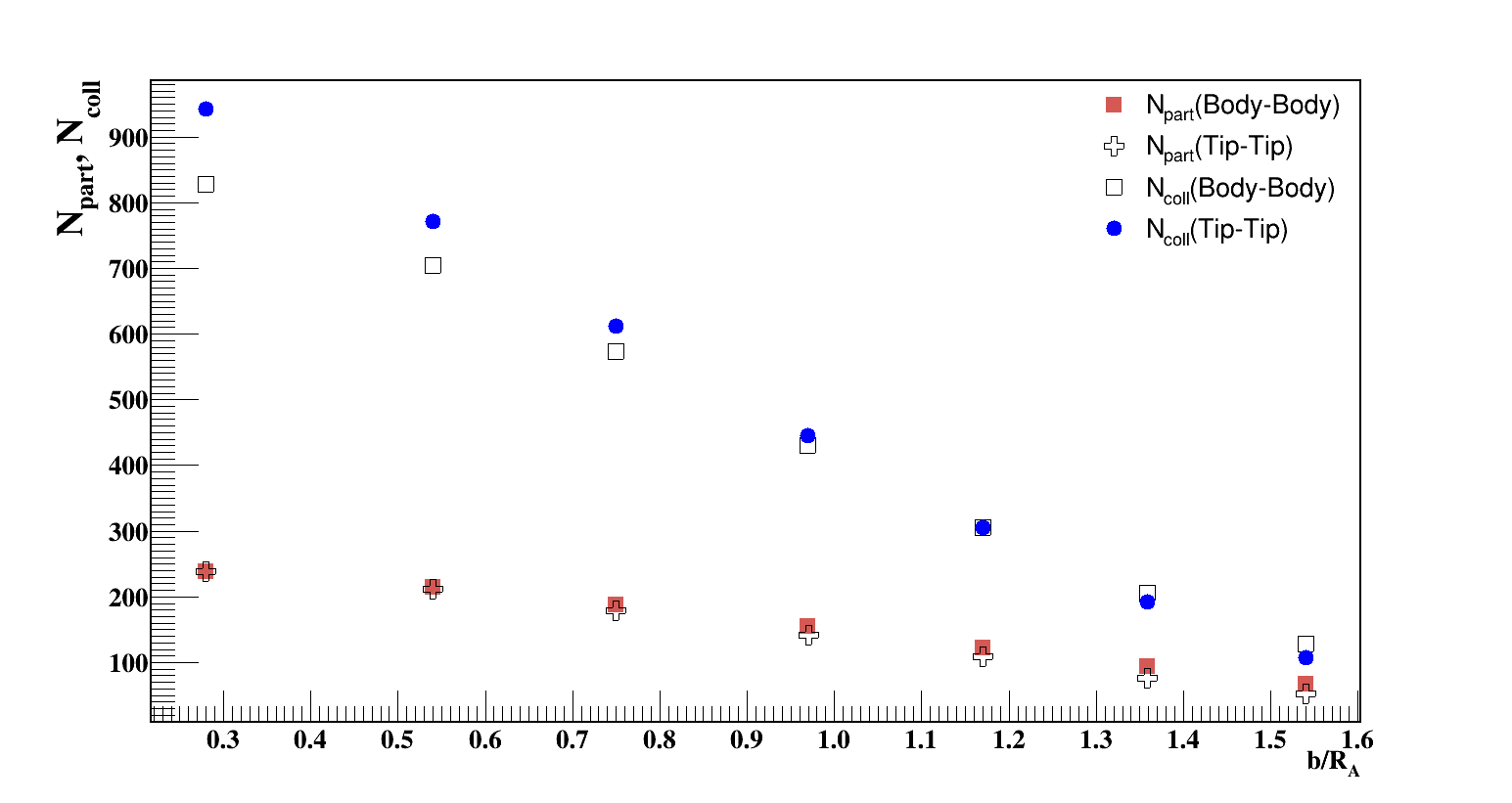}
\caption{Variation of number of participants $N_{part}$ and number of binary collisions $N_{coll}$ with centrality in Xe-Xe collisions.}
\label{npartncoll}
\end{figure}

HYDJET++ is designed to work for spherically symmetric collision systems only. Therefore, the modification was needed to make it work for deformed collision systems \cite{Singh:2017fgm}. The main alteration done is to change the nuclear density profile function. This was a tacky task as HYDJET++ works in cylindrical polar coordinates ($\rho, z, \psi$) unlike AMPT which works in the spherical polar coordinate system ($r, \theta, \phi$). For this, transformation of the deformed Woods-Saxon nuclear density profile function from a spherical polar to a cylindrical polar coordinate system was performed. In spherical polar coordinates, the deformed Woods-Saxon for xenon nucleus is defined as \cite{chaturvedi2016multiplicity}:
\begin{equation}
\rho(r,z,\theta)=\frac{\rho_{0}}{1+\exp\frac{(r-R(1+\beta_{2}Y_{20}+\beta_{4}Y_{40}))}{a}} 
\end{equation} 
where \\
$\rho_{0} =  \rho_{0}^{const} + \text{correction},
\rho_{0}^{const} = \frac{M}{V} =  \frac{3A}{4 \pi R_{A}^3} = \frac{3}{4 \pi R_{l}^3},\\ 
R_{A}=  R(1+\beta_{2} Y_{20} +\beta_{4} Y_{40}),\\
R_{l}=  R_{0}(1+\beta_{2} Y_{20} +\beta_{4} Y_{40}),\\
R=R_{0} A^{1/3},\text{where} \quad  R_{0} = 1.15fm,$ \\
The correction term is calculated as= $\rho_{0}^{const} × (\pi f/R_{A})^{2}$, \\
\text{where} f= 0.54 fm ,\\
$\beta_{2}$ =0.162 \text{and}  $\beta_{4}$ =-0.003 \\
\text{are the deformation parameters \cite{MOLLER20161}},\\
a=\text{diffuseness parameter}= 0.59 fm,\\
$Y_{20} =\sqrt{\frac{5}{16\Pi}}(3\cos^{2}\theta-1)$,\quad \text{and}\\
$Y_{40} = \frac{3}{16\sqrt{\Pi}}(35\cos^{4}\theta-30\cos^{2}\theta + 3)$\\ 
are spherical harmonics.
The geometrical configurations are mainly controlled by $\theta$, integrating all other coordinates over the same range. By changing $\phi$ several other geometries can be obtained but here we will only consider body-body and tip-tip configurations. Performing coordinate transformation we obtained relations $\theta =\tan^{-1} (z/r)$ and $\theta=\tan^{-1} (r/z)$ for body-body and tip-tip geometrical configurations, respectively. Here, r is $\rho$ of cylindrical polar coordinates, avoiding any confusion with that of spherical polar coordinates. To ascertain the model for these two geometrical configurations of Xe-Xe collisions, we have calculated the number of participants ($N_{part}$) and the number of binary collisions ($N_{coll}$) for both geometrical configurations in figure \ref{npartncoll} with respect to the centrality of the event.

\textbf{Soft State:} The soft state of an HYDJET++ event is a thermal hadronic state created on the chemical and thermal freeze-out hypersurfaces derived from the parameterization of relativistic hydrodynamics with preset freeze-out conditions. A thorough description of the physics frameworks of this is available in the equivalent articles \cite{PhysRevC.74.064901, PhysRevC.77.014903}.

The hadronic matter produced in a nuclear collision attains local equilibrium after a short period of time ($<1$fm/c), then expands hydrodynamically. The particle densities at the chemical freeze-out ($T_{ch}$) are so high that the particles cannot be considered free-streaming. Therefore, the presumption of common freeze-outs is not justified. As a result, a more complex scenario, differential freeze-out is ($T_{ch}\geqslant T_{th}$ )is considered. The system expands hydrodynamically with frozen chemical composition in between these two freeze-outs, cools down and finally the hadrons stream out as soon as $T_{th}$ is reached \cite{PhysRevC.73.034905, CLEYMANS2008172, PhysRevC.85.014908, chatterjee2015freeze, AKKELIN2002439}. To generate the initial conditions for chemical and thermal freeze-out hypersurface, the hydrodynamic evolution of this freeze-out hypersurface i.e., the hydrodynamic evolution laws for QCD medium must be calculated. In the soft-state HYDJET++ framework, the QCD medium is assumed to evolve according to the Bjorken boost-invariant hydrodynamics. Therefore, the cooling laws for energy density and temperature are given by \cite{Lokhtin:2000wm}:
\begin{equation}
\epsilon(\tau)\tau^{4/3} = \epsilon_{0}\tau_{0}^{4/3}, \text{and}
\end{equation}
\begin{equation}
T(\tau)\tau^{1/3} = T_{0}\tau_{0}^{1/3}, \text{respectively.}
\end{equation}
Here, $\epsilon_{0}$, and $T_{0}$ are the initial energy density and temperature at the initial proper time $\tau_{0}$ at which the local thermal equilibrium is achieved. The initial energy density at a given impact parameter b is calculated from the following equation \cite{Lokhtin:2000wm}:
\begin{equation}
\epsilon_{0}(b, \tau_{0}) = \epsilon_{0}(b=0, \tau_{0}).\dfrac{T_{AA}(b)}{T_{AA}(0)}.\dfrac{S_{AA}(b)}{S_{AA}(0)}
\end{equation}
where the total initial transverse energy deposition in the mid-rapidity region given by \cite{Lokhtin:2000wm},
\begin{equation}
\epsilon_{0}(b=0, \tau_{0}) = T_{AA}(0).\sigma^{jet}_{NN}(\sqrt{s},p_{0}).\langle p_{T} \rangle ,
\end{equation}
$S_{AA}(b)$ is the effective transverse area of the nuclear overlapping zone at impact parameter b expressed as \cite{Lokhtin:2000wm}:
\begin{equation}
S_{AA}(b) = \int_{0}^{2\pi} d\psi \int_{0}^{r_{max}} r dr
\end{equation}
$\sigma^{jet}_{NN}(\sqrt{s},p_{0}).\langle p_{T} \rangle$ is the first $p_{T}$ moment of the inclusive differential minijet cross-section which is determined by the dynamics of the nucleon-nucleon interactions at the corresponding centre-of-mass energy. $T_{AA}$(b) is the nuclear overlap function calculated using the modified deformed Woods-Saxon nuclear density profile function in cylindrical coordinates $\rho(r,z,\psi)$ as \cite{Lokhtin:2000wm}:
\begin{equation}
T_{AA}(b) = \int_{0}^{\infty} r dr d\psi T_{AA}(r_{1}) T_{AA}(r_{2}) ,
\end{equation}
$T_{A}(r)$ nuclear thickness function is given by-
\begin{equation}
T_{A}(r) = A \int \rho_{A}(r,z,\psi) dz , r_{1,2} = \sqrt{r^{2}+ \dfrac{b^{2}}{4}\pm rb\cos{(\psi)}}
\end{equation}
$r_{1,2}(b,r,\psi)$ are the distances between the centers of colliding nuclei and the jet production vertex M($r\cos \psi,r\sin \psi $) and r is the distance from the nuclear collision axis to M (see figure \ref{jet}). Initial temperature is obtained by a parameterization based on ideal thermal gas approximation  \cite{PhysRevC.85.014908, Andronic:2005yp} where $T_{0}(b=0, \tau_{0})$ and baryon chemical potential $\mu_{0}(b=0, \tau_{0})$ are calculated from the collision energy using the relations:
\begin{equation}
\mu_{0}(b=0, \tau_{0}) = \dfrac{a}{1+b\sqrt{s_{NN}}} ,
\end{equation}
\begin{equation}
T_{0}(b=0, \tau_{0}) = c-d\mu_{B}^{2}-e\mu_{B}^{4}
\end{equation}
Here, a,b,c,d and e are the fitting parameters obtained from the best fit of the particle ratios at various collision energies.
\begin{equation}
T_{0}(b,\tau_{0})= T_{0}(b=0,\tau_{0})\left( \frac{N_{part}(b)}{N_{part}(0)}\right)^{1/3}
\label{T0_npart}
\end{equation}  

\begin{figure}[htbp]
\centering
\includegraphics[scale=0.35]{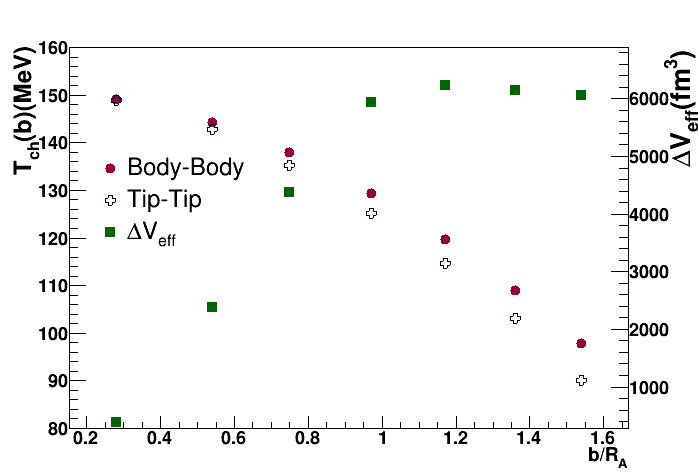}
\caption{Variation of $T_{ch}(b)$ and $\Delta V_{eff}$ with centrality in Xe-Xe collisions at 5.44 TeV.}
\label{tch_veff}
\end{figure}

\begin{figure}[htbp]
\centering
\includegraphics[scale=0.22]{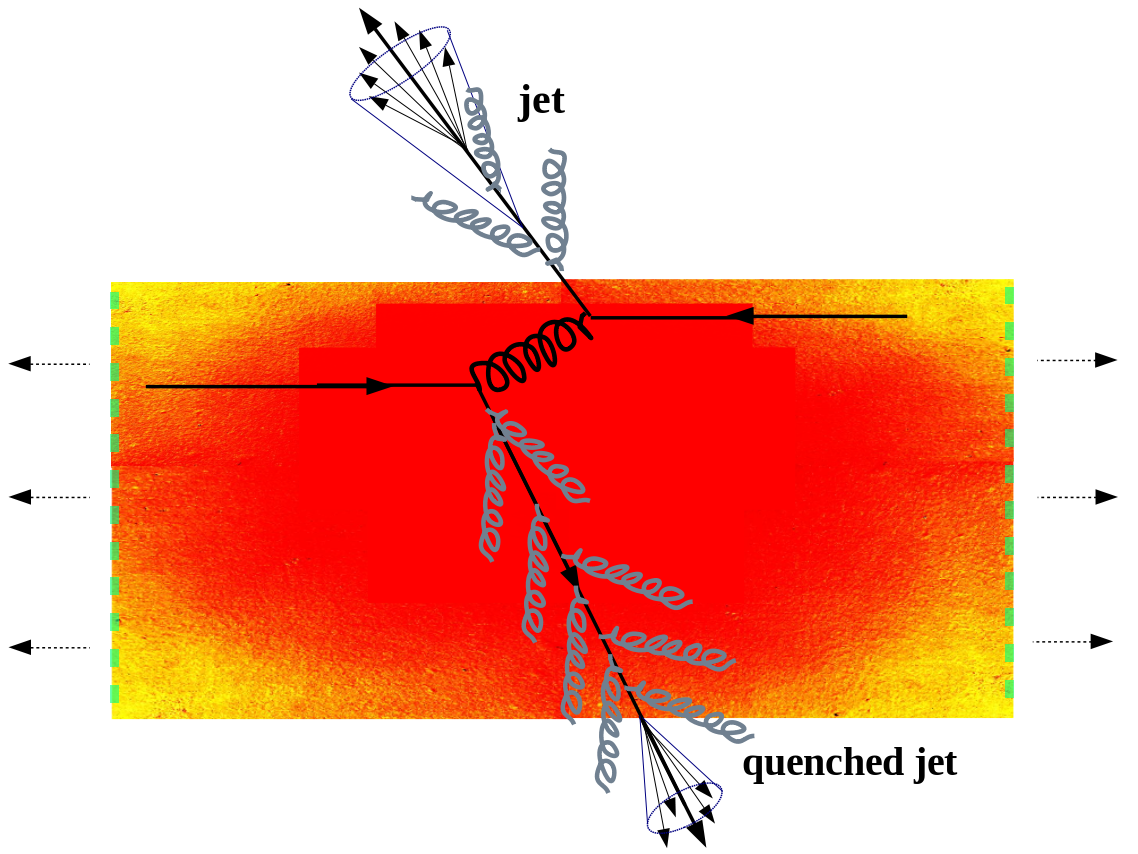}
\caption{A schematic illustration of the high-$p_{T}$ parton traversing the QGP medium in Heavy Ion Collision \cite{Tachibana:2014yai}.}
\label{j_jq}
\end{figure}

The temperature for higher collision centralities is obtained by converting fixed freeze-out hypersurface into centrality (or N$_{part}$) dependent hypersurface, a modification implemented in the soft particle production of HYDJET++ (equation \ref{T0_npart}). $\mu_{B}$ is centrality independent as the value of baryon chemical potential is zero at LHC energies. Hadron multiplicities are obtained via effective thermal volume approximation and Poisson multiplicity distribution around its mean value, which is proportional to the number of participating nucleons at a given impact parameter. The change in effective thermal volume between body-body and tip-tip configuration and variation of chemical freeze-out temperature with respect to collision centrality for body-body and tip-tip configuration of Xe+Xe collision at 5.44 TeV have been presented in figure \ref{tch_veff}.

\section{Jet and Jet Quenching in HYDJET++}
\label{j and jq}
\vspace{-2ex}

Precisely, jets are collimated spray of particles corresponding to hard-scattered partons. The investigation of jet production and fragmentation provides a test to understand perturbative and non-perturbative QCD. Jets produced in initial stage of heavy-ion collisions, probe the early hot and dense quark gluon plasma phase of the fireball evolution. These interact with the medium resulting in added induced radiation which emits at smaller angles inside the jet cone leading to broadening of the jet profile, thereby modifying the high-$p_{T}$ particle spectra \cite{Tachibana:2014yai}. This phenomenon is called as Jet quenching and provides an insight of the QGP properties in heavy-ion collisions (HIC) (see figure \ref{j_jq}). It is one of the smoking-gun signatures of QGP which exemplifies the microscopic degrees of freedom that emerged as QCD became deconfined.

\begin{figure*}[htbp]
\centering
\includegraphics[scale=0.32]{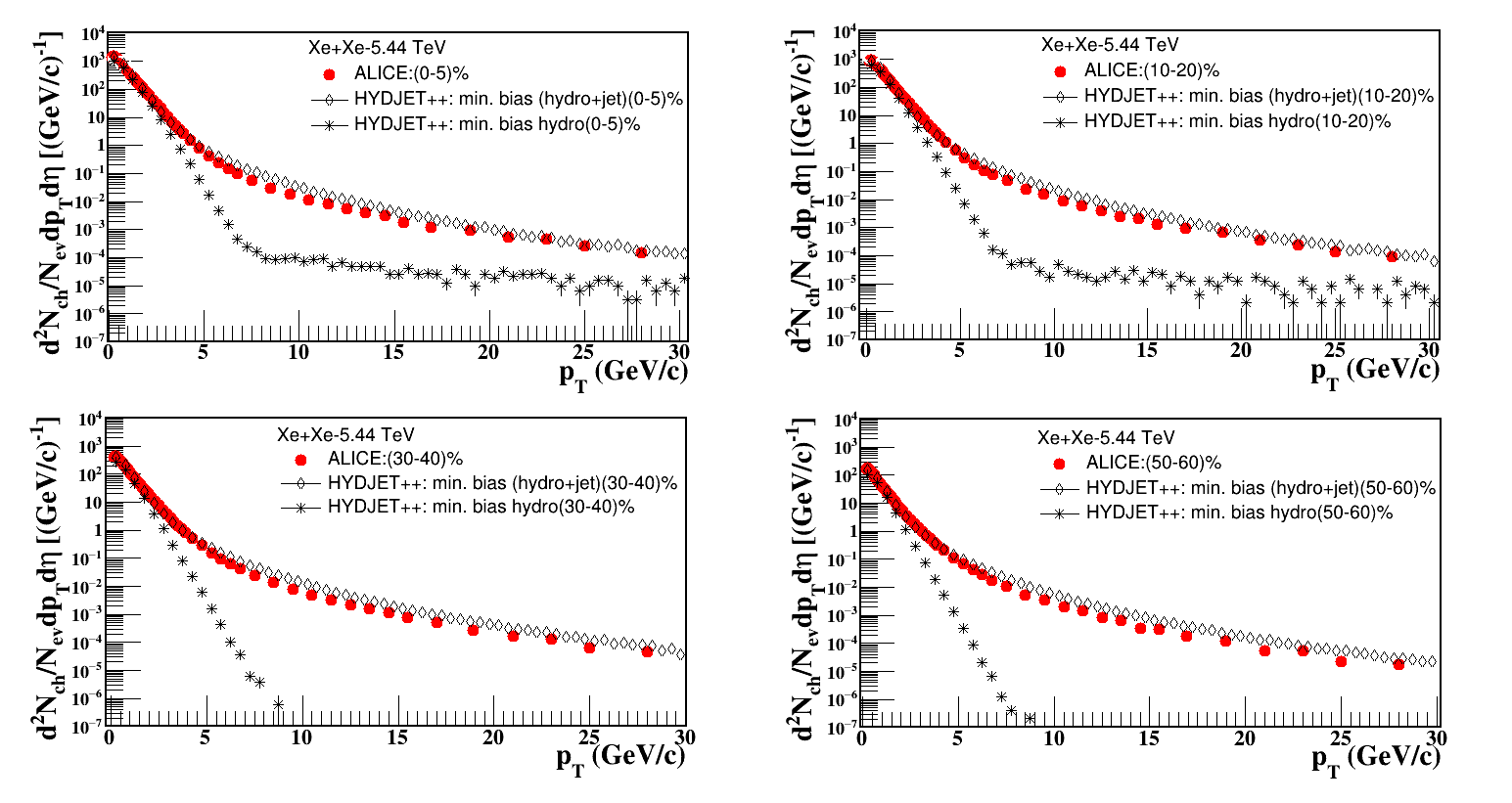}
\caption{Transverse momentum spectra of all charged particles with and without jet part for minimum bias collisions over centrality along with ALICE experimental data for comparison \cite{ALICE:2018hza}. Here we have used $|\eta|<0.8$ pseudorapidity cut.}
\label{pt-spectra}
\end{figure*}

Jet quenching in ultra-relativistic heavy-ion collisions, involves two different medium-induced mechanisms for jet energy loss: elastic collisions with medium constituents and induced bremsstrahlung processes. Scatterings between hard partonic jets with medium constituents induce additional radiation which rips apart parent parton's energy. This is called as radiative energy loss \cite{gyulassy1994multiple, wang1995landau, baier1995induced, Baier:1996kr, BAIER2003209c, baier1998radiative}. Another is the collisional energy loss occuring in binary elastic collisions \cite{Bjorken:1982tu, Mrowczynski:1991da, thoma1991collisional, Lokhtin:1997vm}. The  collisional energy losses represents an incoherent sum over all rescatterings and is almost independent of the initial parton energy. The radiative energy losses of a high energy parton dominate over the collisional losses almost by an order of magnitude. The amount of energy loss varies linearly with path-length in case of collisonal energy loss while the dependence is quadratic for radiative energy loss due to interference effects.

\begin{figure*}[htbp]
\centering
\includegraphics[scale=0.32]{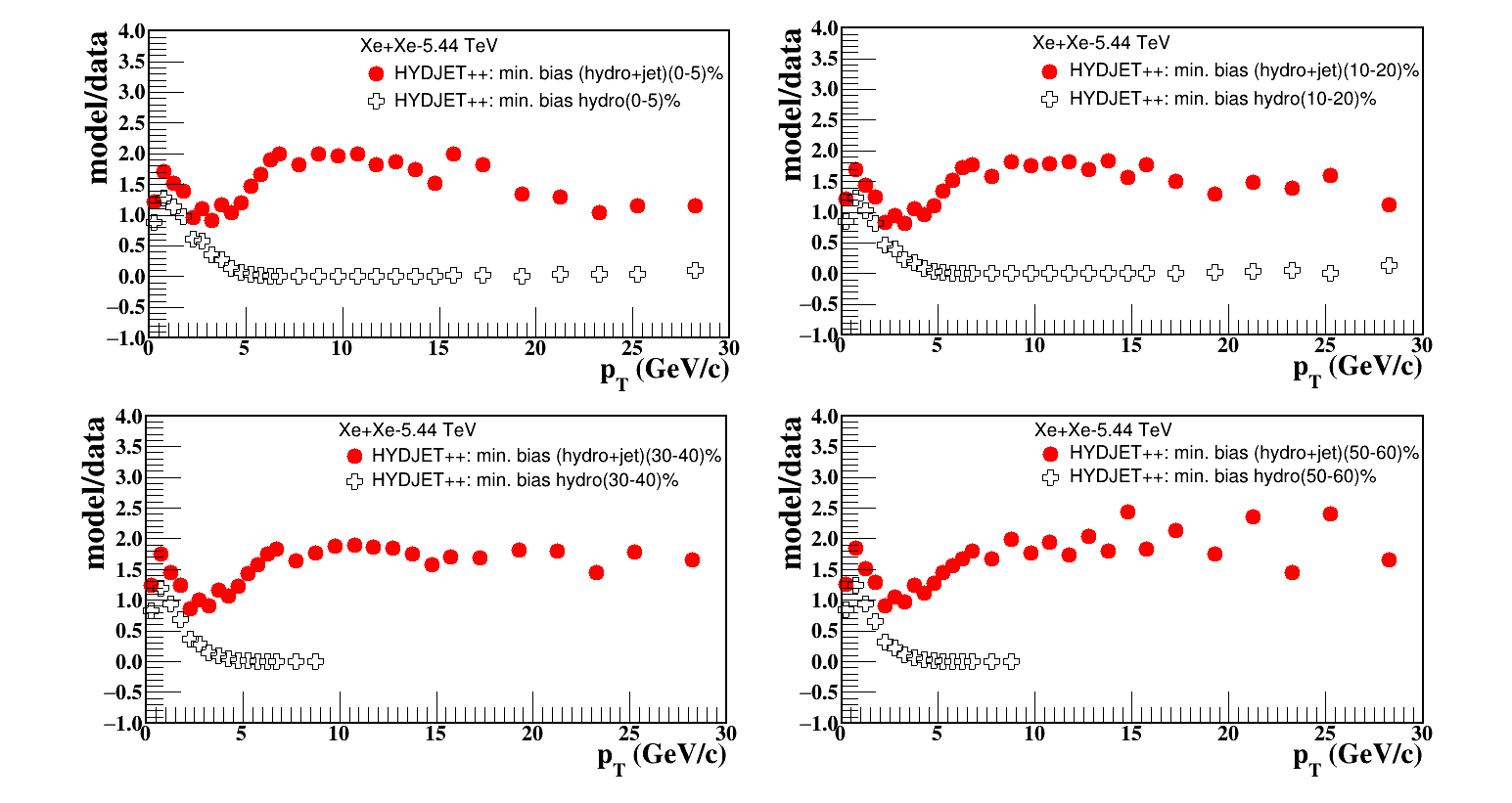}
\caption{Ratio of Model results to ALICE experimental data \cite{ALICE:2018hza} for transverse momentum spectra of all charged particles with and without jet part for minimum bias collisions over centrality. Here we have used $|\eta|<0.8$ pseudorapidity cut.}
\label{pt-mbyd}
\end{figure*}

The suppression of particle spectra in A+A collisions is measured by the nuclear modification factor $R_{AA}$ . For a given centrality bin $\Delta c$ (determined experimentally via charged hadron multiplicities) $R_{AA}$ is defined as the ratio of particle yield in A+A collisions to that in p+p collisions scaled with the number of binary collisions as follows \cite{Zakharov:2021uza}:-
\begin{equation}
R_{AA} = \dfrac{d^{2}N_{AA}/d^{2}p_{T}dy}{\langle N_{coll} \rangle_{\Delta c} d^{2}N_{pp}/d^{2}p_{T}dy},
\end{equation}

where, $\langle N_{coll}\rangle _{\Delta c}$ is the number of binary collisions for centrality class $\Delta c$.

The relation between centrality c and the impact parameter b of A+A collision is calculated according to the Glauber model. Theoretical calculations of the nuclear suppression are performed assuming that the medium jet modification occurs only in A+A collisions, and absent in p+p collisions, i.e., the p+p yield is calculated in the standard pQCD approach. Here, we have taken p+p from ALICE experiment. Measured $R_{AA}$ if smaller than 1 indicates strong suppression of jets.

\begin{figure*}[htbp]
\centering
\includegraphics[scale=0.32]{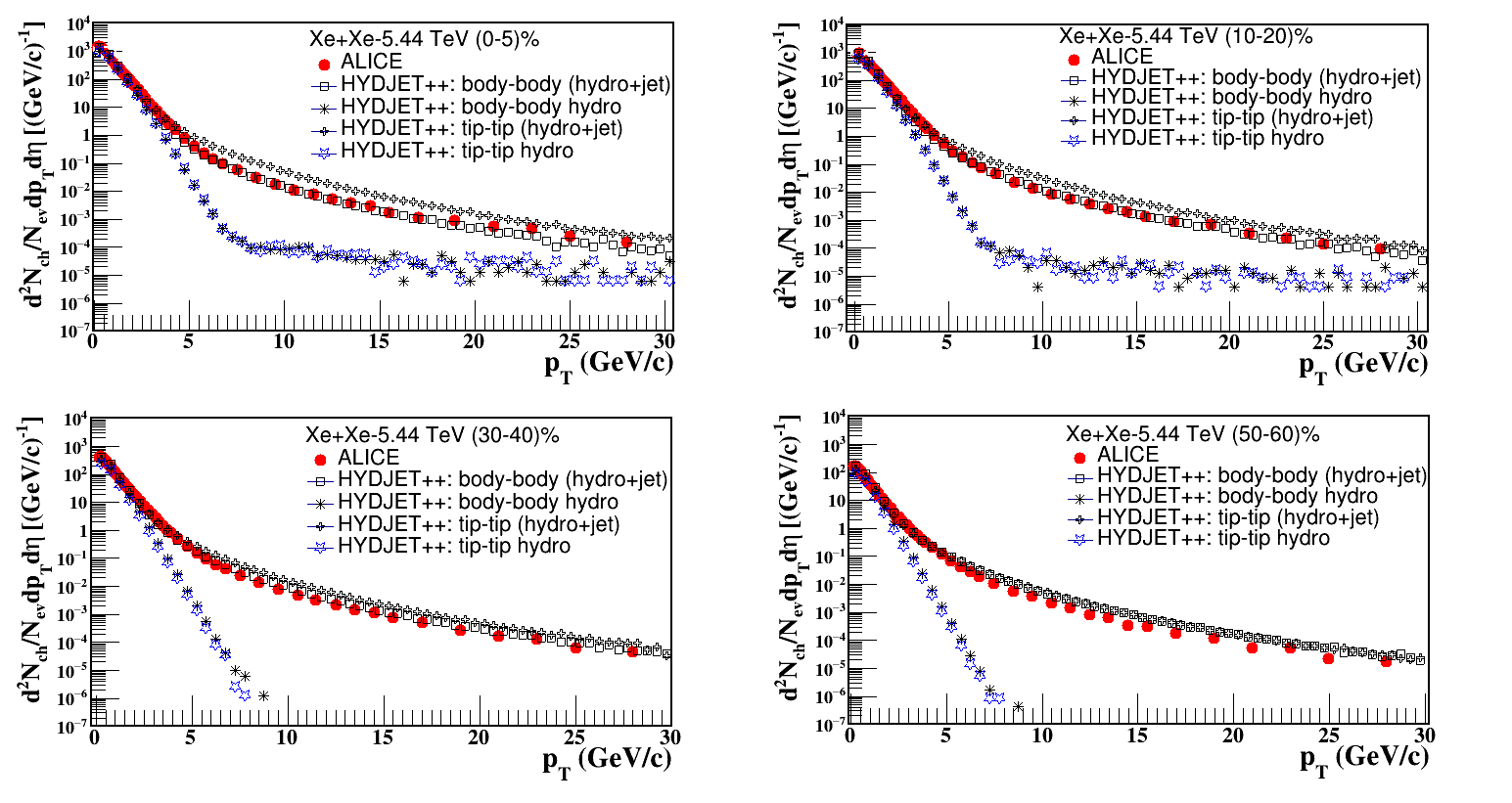}
\caption{Transverse momentum spectra of all charged particles with and without jet part for body-body and tip-tip collisions over centrality along with ALICE experimental data for comparison \cite{ALICE:2018hza}. Here we have used $|\eta|<0.8$ pseudorapidity cut.}
\label{pt-spectra-configs}
\end{figure*}

\section{Results and Discussions}
\label{results}
\vspace{-2ex}
\subsection*{A. Transverse momentum spectra}

\begin{figure*}[htbp]
\centering
\includegraphics[scale=0.32]{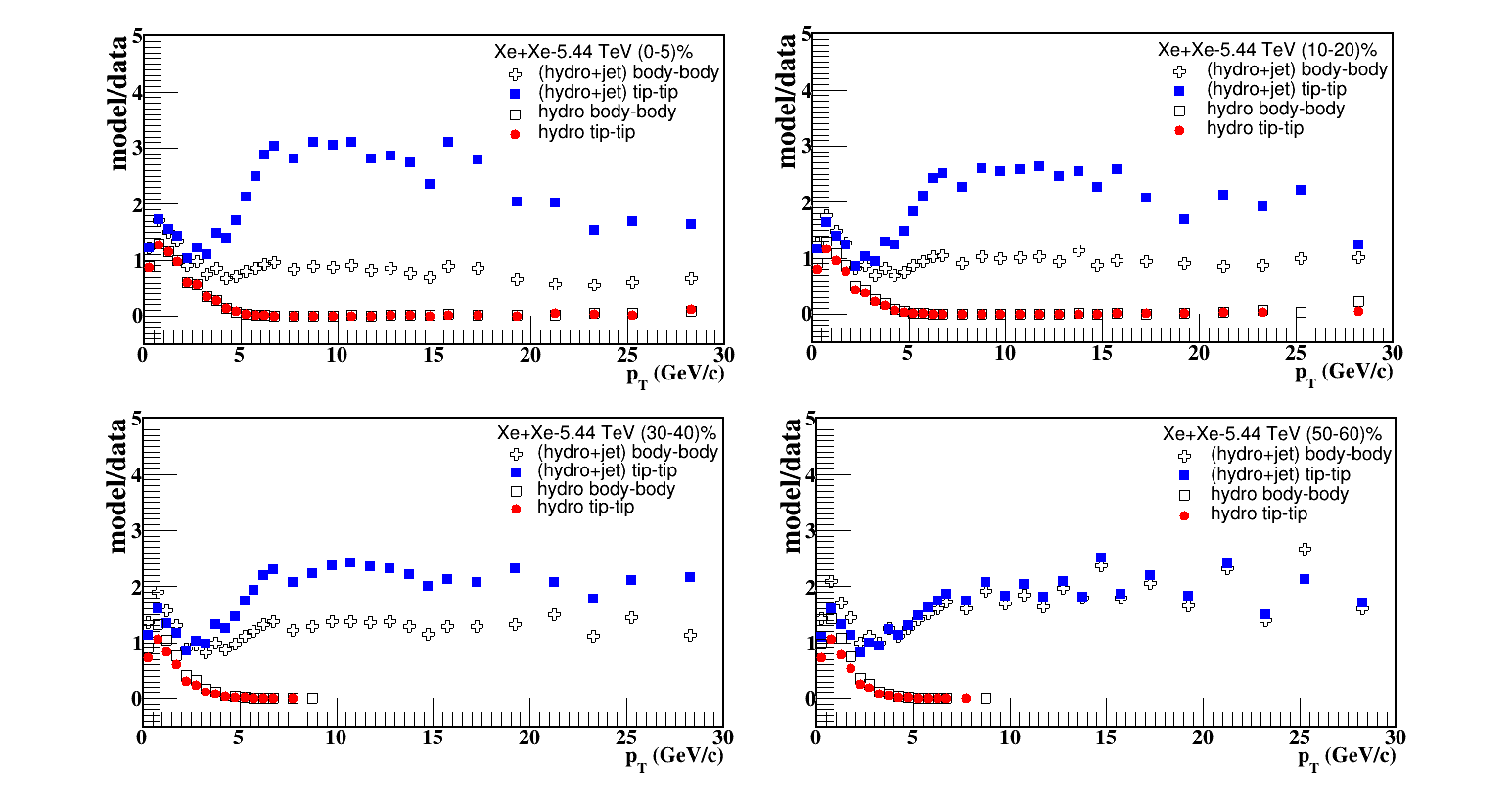}
\caption{Ratio of Model results to ALICE experimental data \cite{ALICE:2018hza} for transverse momentum spectra of all charged particles with and without jet part for body-body and tip-tip collisions over centrality. Here we have used $|\eta|<0.8$ pseudorapidity cut.}
\label{pt-mbyd-configs}
\end{figure*}

\begin{figure}[htbp]
\centering
\includegraphics[width=9.5cm,height=5.5cm]{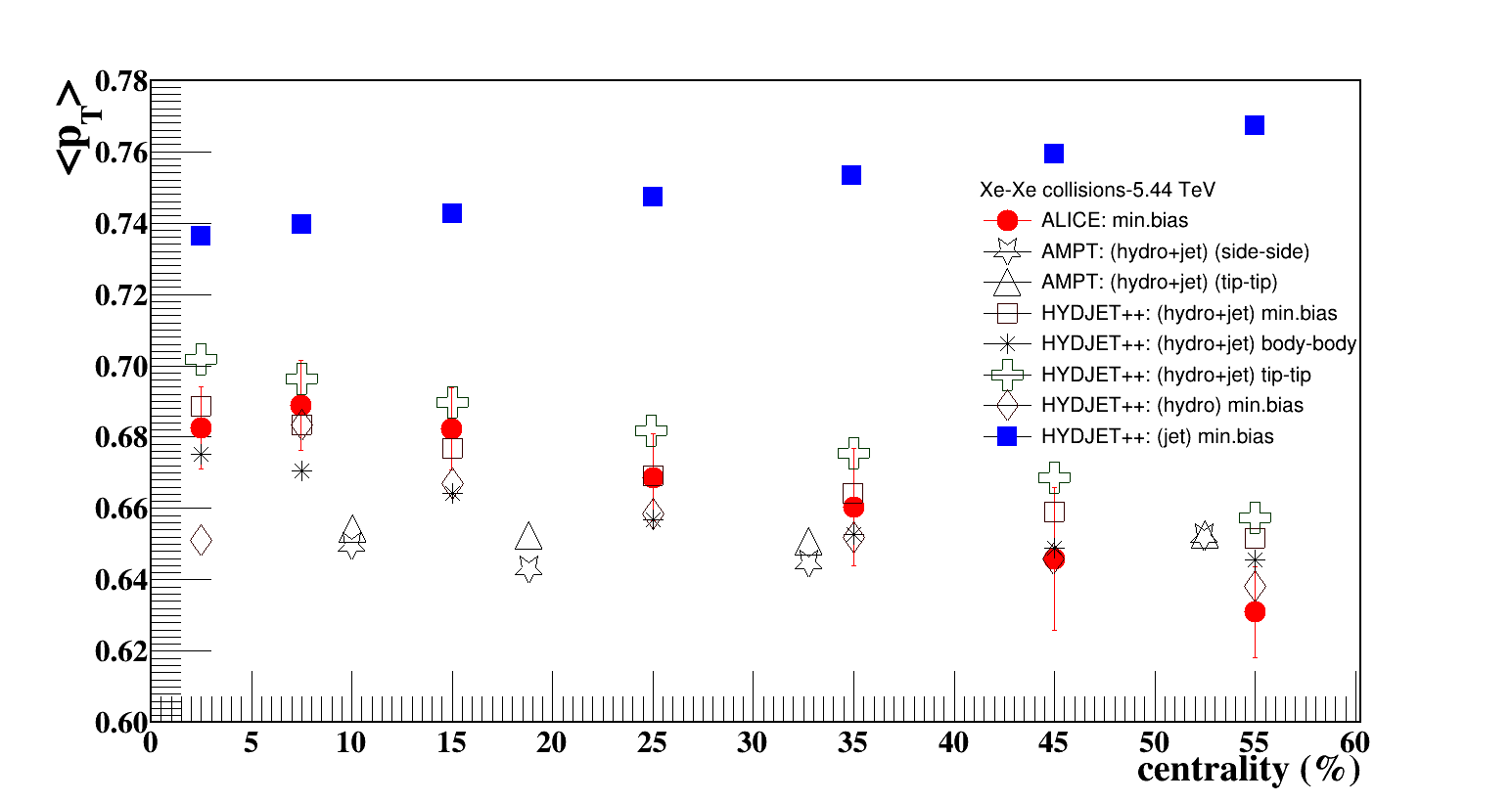}
\caption{Variation of average transverse momentum $\langle p_{T} \rangle$ with respect to collision centrality at $|\eta|<0.8$ for minimum-bias, body-body and tip-tip collisions \cite{PhysRevC.103.014903}. The results show average $p_{T}$ for hydro, jet and (hydro+jet) part separately comparing them with ALICE experimental data \cite{ALICE:2018hza} and AMPT model results in string melting version \cite{kundu2019study}.}
\label{avg-pt}
\end{figure}

Transverse momentum spectra of charged hadrons is a well-studied observable for sure that furnish vital information about the high-density deconfined state of strongly-interacting quark-gluon plasma. In figure \ref{pt-spectra}, we present minimum bias $p_{T}$-spectra of charged hadrons with and without the contribution of jet in Xe-Xe collisions at 5.44 TeV via HYDJET++ model \cite{PhysRevC.103.014903}. The results have been conducted in most-central, semi-central, semi-peripheral and most peripheral classes of collisions and compared with ALICE experimental data. We find a suitable match of model $p_{T}$-spectra including jet part with ALICE experimental results \cite{ALICE:2018hza} while without jet part, model completely underpredicts the experimental data. Below 3.0 GeV/c, the collective flow hydrodynamics dominates the spectra but beyond that, hard scatterings come into play and hence the spectra is obtained. As a function of collision centrality, $p_{T}$ spectra decreases as we move from most-central to most-peripheral results. The slope of $p_{T}$-distribution being inversely proportional to the fireball temperature indicates that central collisions have higher fireball temperature than in peripheral collisions, as expected \cite{PhysRevC.103.014903}.

The ratio of HYDJET++ model results to ALICE experimental data \cite{ALICE:2018hza} shown in figure \ref{pt-mbyd} for transverse momentum spectra in minimum bias xenon-xenon collisions is around 1.8 for hydro+jet part while without contribution from jet or hard part, the ratio is less than 1.0, almost two times petite, decreases and eventually goes to 0. Visualising these in body-body and tip-tip configurations' scenario is performed in our next plot (figure \ref{pt-spectra-configs}) where  $p_{T}$-spectra for body-body and tip-tip collisions with and without contribution from jet part is presented as a function of $p_{T}$ and collision centrality. Total (hydro+jet) $p_{T}$-spectra of tip-tip collisions is higher than body-body collisions where body-body collisions resemble more the ALICE experimental data in comparison to tip-tip collisions. Without jet or hard part, the transverse momentum distribution is very small and completely underestimates the experimental data. We can also clearly see that, without jet contribution the $p_{T}$-distribution is same for both the geometrical configurations while the effect of geometry becomes visible only when the contribution of hard scatterings (jet part) is included in the total $p_{T}$-distribution. However, at low $p_{T}$ ($p_{T}<3.0$ GeV/c) there is is no difference between hydro and jet, body-body and tip-tip collisions. This is evident from figure \ref{pt-mbyd-configs} where ratio of HYDJET++ model to ALICE experimental data results \cite{ALICE:2018hza} have been presented. The difference between the two geometrical configurations is observed in total (hydro+jet) $p_{T}$-spectra being maximum in most-central collisions and decreasing with increasing centrality while it is absent in hydro $p_{T}$-spectra.  Here, the tip-tip ratio is almost 3 times higher than the ratio for body-body collisions. Also, as a function of collision centrality, the difference in the two geometrical configurations vanishes. This is due to the fact that the effect of deformation falls as we move from central to peripheral collisions.

\begin{figure*}[htbp]
\centering
\includegraphics[scale=0.32]{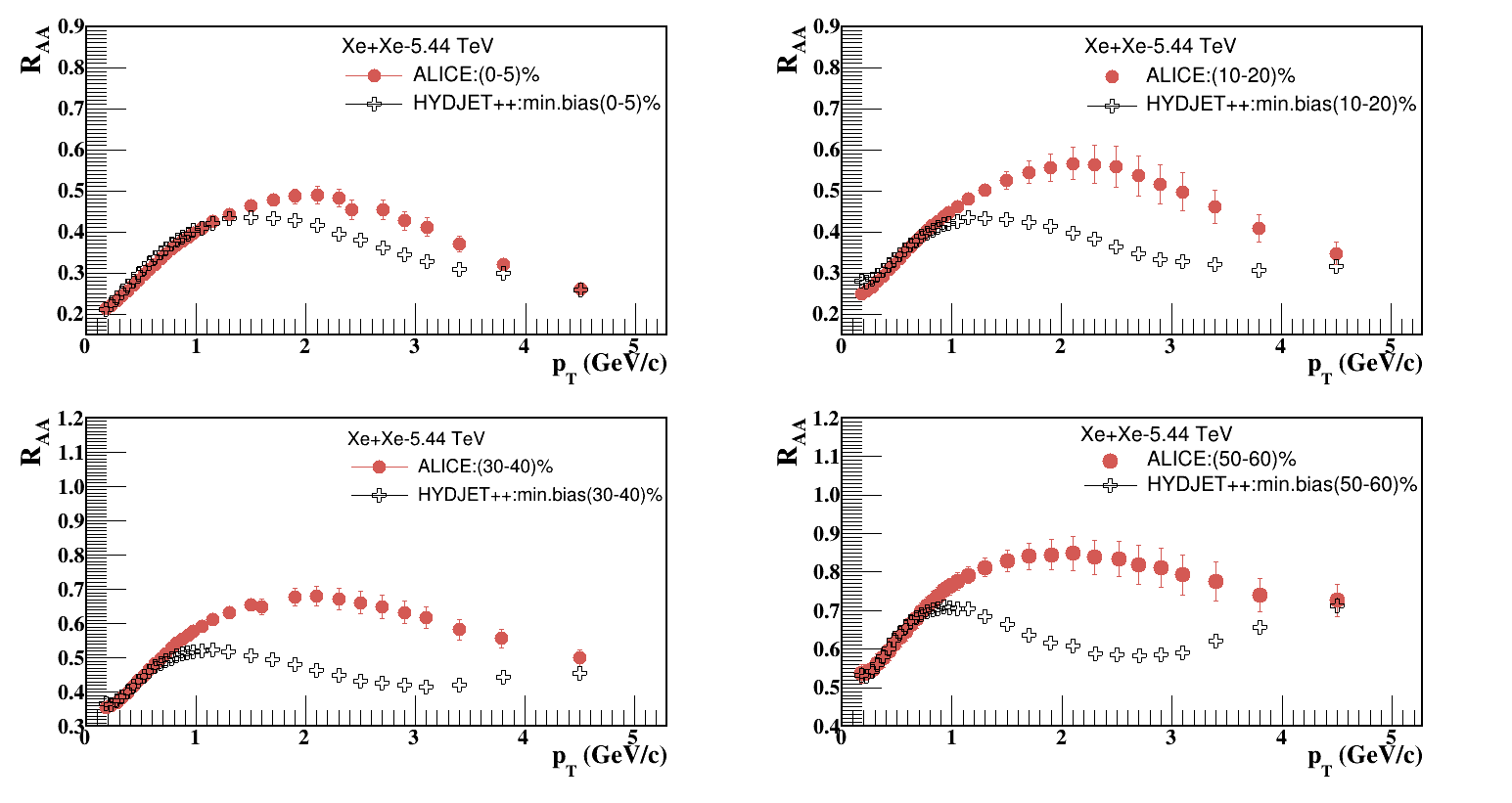}
\caption{Nuclear Modification Factor $R_{AA}$ of charged hadrons with respect to $p_{T}$ for minimum bias, body-body and tip-tip collisions over centrality along with ALICE experimental data for comparison \cite{ALICE:2018hza}. Here we have used $|\eta|<0.8$ pseudorapidity cut.}
\label{raa-pt}
\end{figure*}

In figure \ref{avg-pt}, we present minimum bias average transverse momentum ($\langle p_{T} \rangle$) spectra as a function of collision centrality for total (hydro+jet), only hydro, and only jet part of Xe-Xe collisions at 5.44 TeV. The model results have been compared with ALICE experimental data \cite{ALICE:2018hza} and those in body-body and tip-tip geometrical configurations. We observe that minimum bias $\langle p_{T} \rangle$ for hydro+jet part decreases as we move from most-central to most-peripheral class of collisions\cite{PhysRevC.103.014903}. Similar is observed for hydro part whereas for jet part $\langle p_{T} \rangle$ weakly rises (almost negligible) with impact parameter. Minimum bias HYDJET++ results for total and hydro part show suitable match with ALICE experimental data whereas jet part shows opposite behaviour. An increase of $\langle p_{T} \rangle$ with collision centrality visible in collision systems indicates increasing transverse radial flow. Similar behaviour are inferred from body-body and tip-tip collision results. Also, at a particular collision centrality, $\langle p_{T} \rangle$ is higher for tip-tip collisions than for body-body collisions. We have also compared HYDJET++ results with those from AMPT model where we find that, $\langle p_{T} \rangle$ from AMPT model \cite{kundu2019study} in string melting version shows a weak centrality dependence. Here, the qualitative behaviour is very similar to those from jet part of HYDJET++ model results although underestimating quantitatively. 

\subsection*{B. Nuclear Modification Factor $R_{AA}$}

\begin{figure}[htbp]
\centering
\includegraphics[width=8.0cm,height=6.5cm]{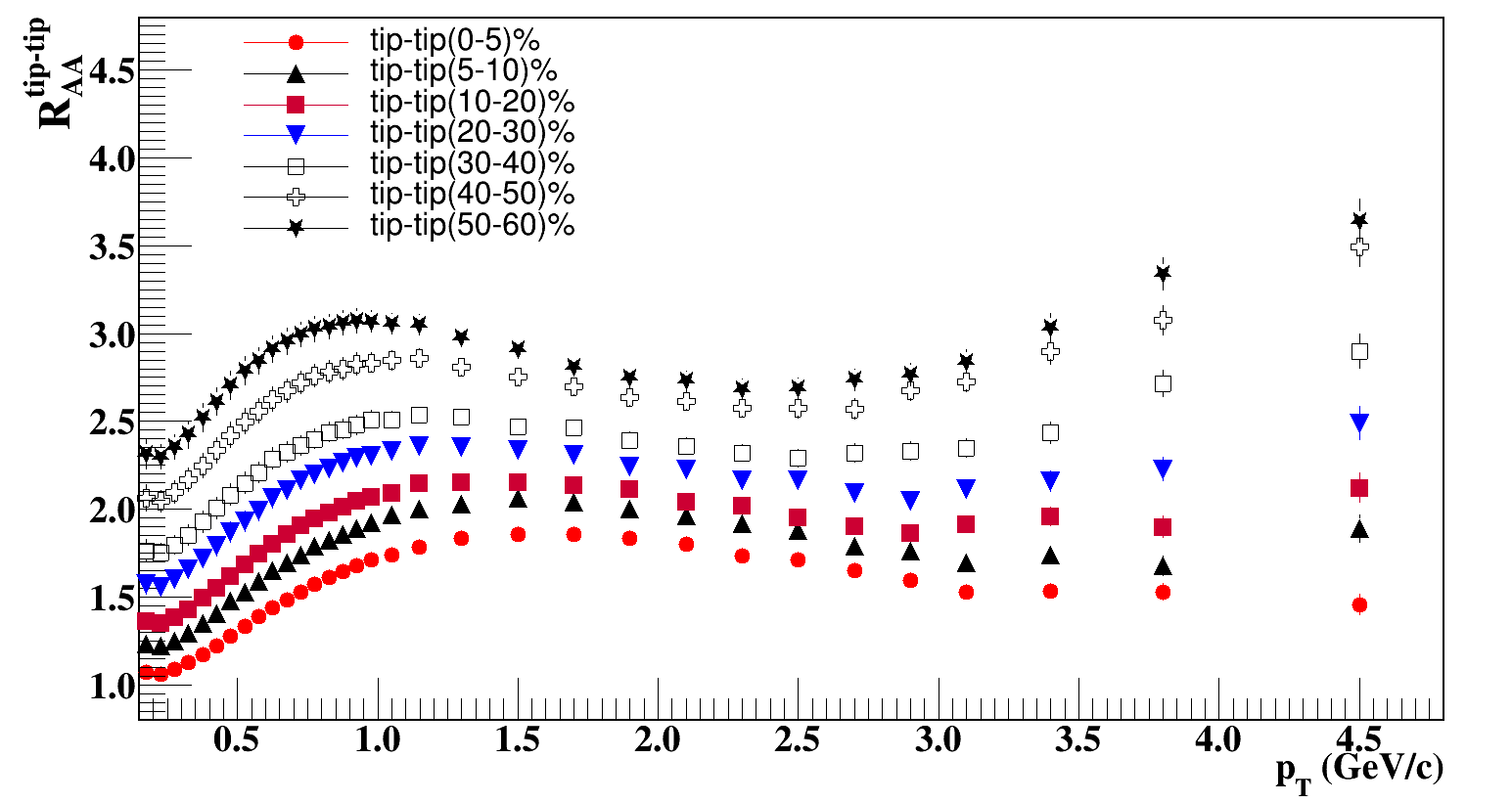} 
\caption{Nuclear modification factor $R_{AA}$ of charged hadrons with respect to $p_{T}$ in tip-tip configuration over seven classes of centralities.}
\label{raa-pt-tt}
\end{figure}

Nuclear Modification Factor $R_{AA}$ is a powerful tool which can be used for calibrating parton-medium interaction mechanism, independently from the medium evolution. In order to determine nuclear modification factor $R_{AA}$ for primary charged hadrons in deformed Xe-Xe collisions at 5.44 TeV, an interpolated proton-proton (pp) reference spectrum for the same energy scaled by the number of binary collisions $\langle N_{coll} \rangle$ is used. The resulting $R_{AA}$ is studied as a function of transverse momentum and collision centrality (in figure \ref{raa-pt}) from most-central to most-peripheral class of collisions and compared to ALICE experimental data \cite{ALICE:2018hza}. Minimum bias nuclear modification factor $R_{AA}$ from HYDJET++ model show a suitable match with ALICE experimental results upto $\sim 1.0$ GeV/c. The suppression increases at low $p_{T}$, reaches maximum at around $p_{T}\simeq 2$ GeV/c. This is caused by the interplay of soft and hard processes. In the range $2 \leq p_{T} \leq 5$  GeV/c, $R_{AA}$ falls reaching minimum. HYDJET++ presents similar characteristic behaviour as experiment. However, it underestimates experimental data above 2 GeV/c. As a function of collision centrality, nuclear modification factor $R_{AA}$ increases with collision centrality and then decreases in most-peripheral collisions. 
\begin{figure}[htbp]
\centering
\includegraphics[width=8.8cm,height=6.3cm]{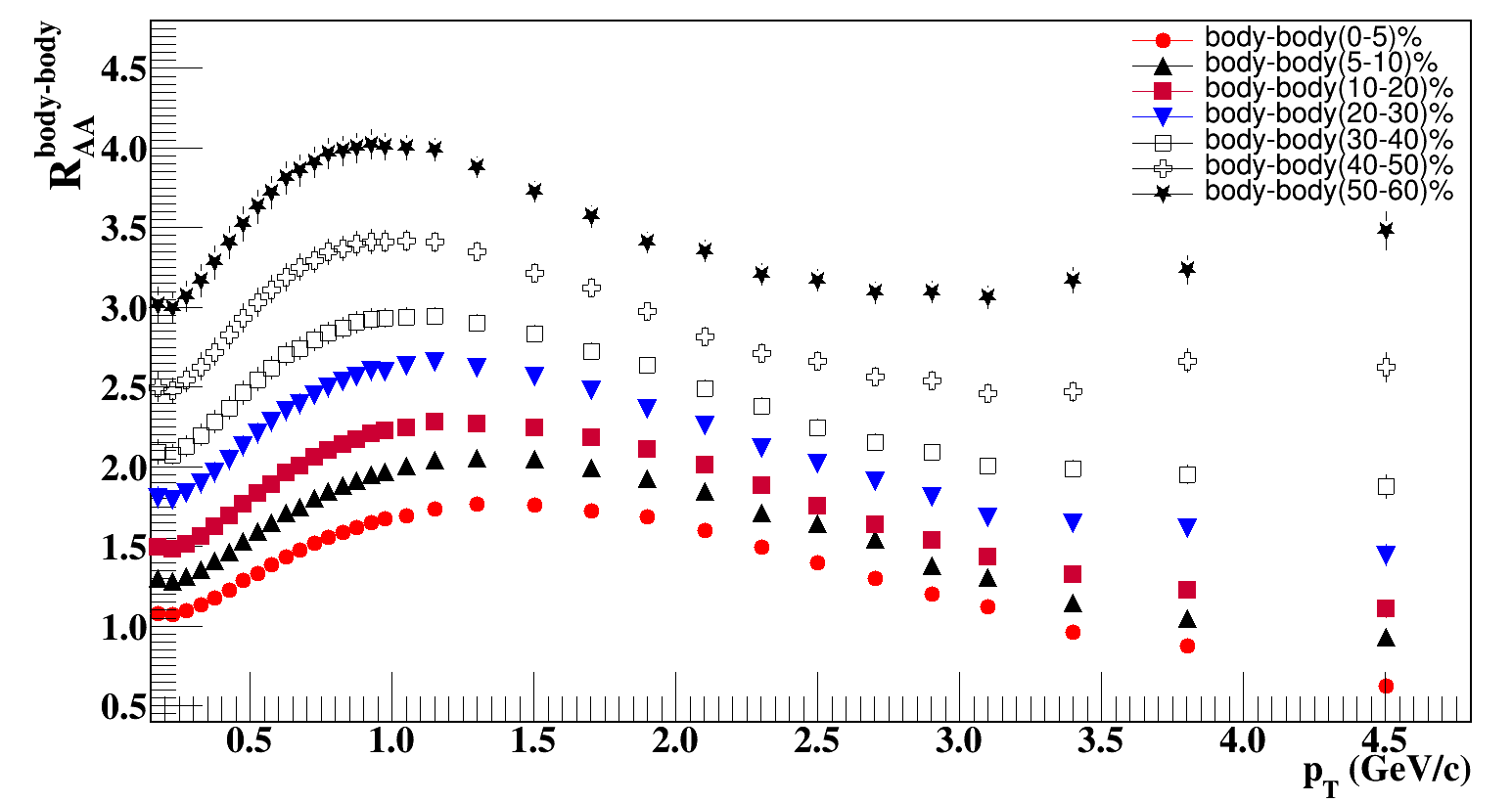} 
\caption{Nuclear modification factor $R_{AA}$ of charged hadrons with respect to $p_{T}$ in body-body configuration over seven classes of centralities.}
\label{raa-pt-bb}
\end{figure}
Centrality dependence of nuclear modification factor $R_{AA}$ in body-body and tip-tip geometrical configurations can be more clearly seen in figures \ref{raa-pt-bb} and \ref{raa-pt-tt} where we see that $R_{AA}$ increases as we move from most-central to most-peripheral class of collisions. As a function of transverse momentum, the behavior is attributed to the fact that in HYDJET++ medium is treated using a scaling solution obtained by Bjorken hydrodynamics. Outcomes have been shown from most central (0-5)\% class of collisions to most peripheral (50-60)\% class of collisions. Each distribution of $R_{AA}$ at a particular centrality of the collision shows a zenith at $\sim$1.5 GeV/c in most central collisions and this peak value shifts to  $\sim$1.0 GeV/c as we reach most peripheral collisions. The nuclear modification factor is higher in body-body collisions being $\approx$1.32 times higher than tip-tip collisions. The peak value rises almost 2.22 times for body-body and 1.77 times for tip-tip geometrical configurations as we move from the most central to most peripheral collisions.

The transverse yield of tip-tip collisions is higher than the transverse yield of body-body collisions. However, the anisotropic flow of body-body collisions is higher than tip-tip collisions. The fireball medium produced in tip-tip collisions has a somewhat isotropic shape in the transverse plane (R$_{x}$=R$_{y}$), the anisotropy generated mostly by the fluctuations in the medium. On the other hand, body-body collisions produce a medium having some elliptic deformation which is also contributed by the quadrupole deformation of the colliding nuclei. As a result, the medium in body-body collisions is less viscous than the medium in tip-tip collisions. Thus, despite the yield is higher in tip-tip collisions, the formation of the fireball medium in such geometrical configuration results in a smaller quenching of the high-$p_{T}$ particle spectra, and hence smaller $R_{AA}$, compared to body-body collisions.

\subsection*{C. Suppression in terms of $R_{CP}$}
We here in our study have not calculated $p_{T}$-spectra for p+p collisions at 5.44 TeV center-of-mass energy using HYDJET++ model, reason being the physical restrictions of the model as it is only applicable for symmetric AA collisions of heavy ($A\geq40$) ions at high energies\cite{refId0, PhysRevC.103.034903}. Therefore, the suppression is studied in terms of $R_{CP}$ which is defined as the ratio of the modification in charged particle $p_{T}$-spectra at a given collision centrality to the modification in $p_{T}$-spectra in peripheral collisions. It is expressed as \cite {PhysRevC.103.034903}:

\begin{equation}
R_{CP} =  \dfrac{\left(\dfrac{d^{2}N_{ch}}{N_{ev}dp_{T}d\eta}\right)^{central}}{\left(\dfrac{d^{2}N_{ch}}{N_{ev}dp_{T}d\eta}\right)^{peripheral}},
\end{equation}

\begin{figure}[htbp]
\centering
\includegraphics[width=8.0cm,height=6.5cm]{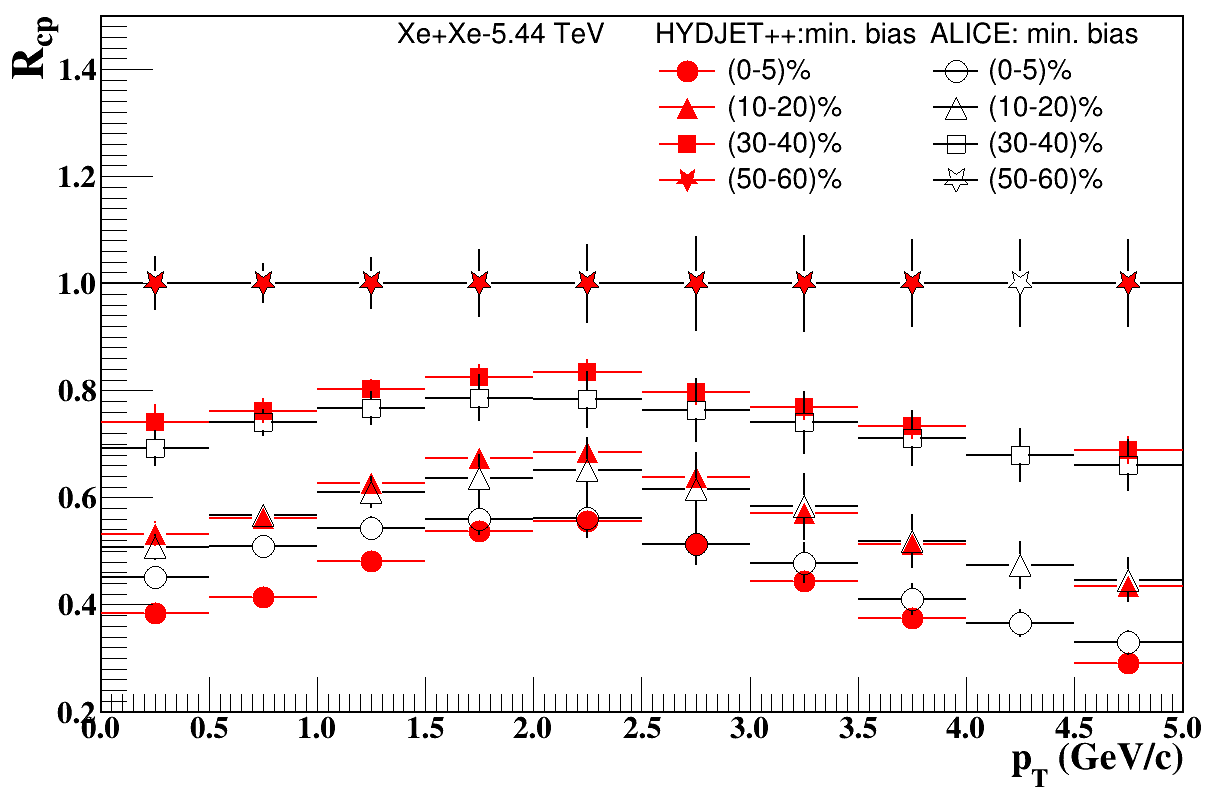} 
\caption{Suppression in terms of $R_{CP}$ of charged hadrons with respect to $p_{T}$ in minimum bias collisions over various classes of centralities.}
\label{rcp-pt}
\end{figure}
\begin{figure}[htbp]
\centering
\includegraphics[width=8.0cm,height=6.5cm]{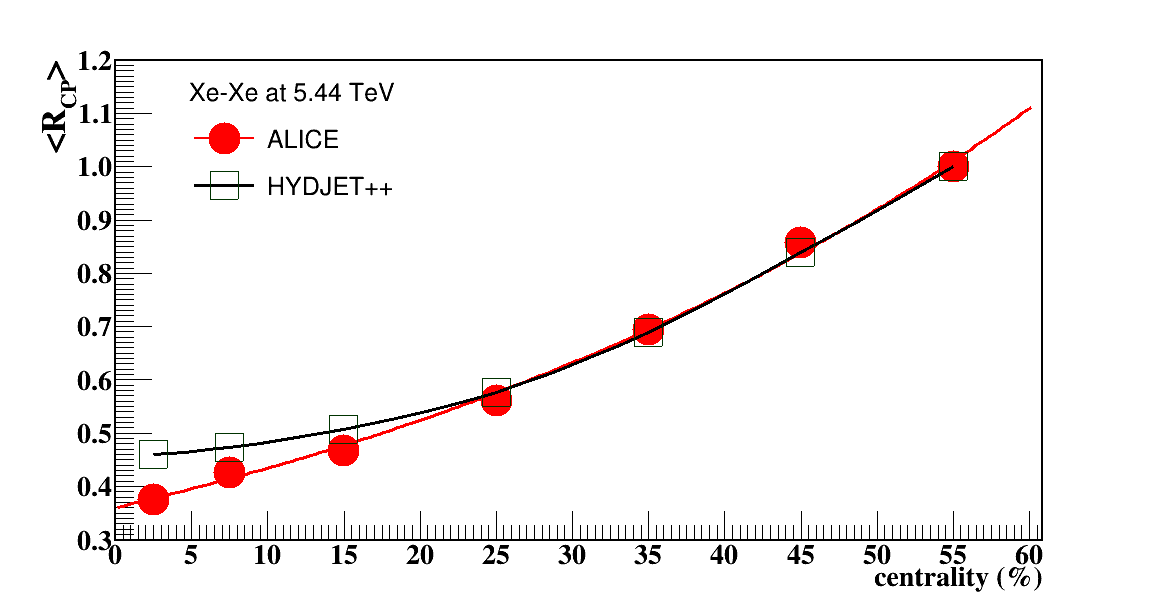} 
\caption{Average relative suppression in terms of $ \langle R_{CP} \rangle $ of charged hadrons in minimum-bias Xe-Xe collisions as a function of collision centrality along with ALICE experimental data for comparison \cite{ALICE:2018hza}.}
\label{avg_rcp}
\end{figure}


dividing each of them by their corresponding average number of the binary collisions $\langle N_{coll}\rangle$. In order to present better comparison of $R_{CP}$ values as a function of $\langle N_{part}\rangle$ or centrality, we chose reference collision system to be the same for all systems ($\langle N_{part}\rangle \geq 30$). Using this technique we have estimated $R_{CP}$ in figure \ref{rcp-pt}. We have compared our results with ALICE experimental data \cite{ALICE:2018hza}. As a function of transverse momentum, the relative suppression $R_{CP}$ increases as we move from low transverse momenta, finds a maximum at around 2.25 GeV/c and decreases as we move towards higher $p_{T}$ region. HYDJET++ model results suitably match with ALICE experimental data \cite{ALICE:2018hza} within error bars except on the very low $p_{T}$ region. As a function of collision centrality, $R_{CP}$ increases as we move from most-central collisions and then saturates (maximum) in most-peripheral collisions. This can be evidently seen in figure \ref{avg_rcp} where average relative suppression in terms of $R_{CP}$ is plotted as a function of collision centrality. $\langle R_{CP} \rangle$ shows a positive correlation with centrality. Model outcomes overpredict ALICE experimental data in central collisions while beyond 30\% collisions centrality we find a suitable agreement with the ALICE experimental data.

\section{Summary and Outlook}
\label{summary}
In a succinct way of our work, we have tried to make a conscientious work on deformed Xenon-Xenon collisions at $\sqrt{s_{NN}}$ = 5.44 TeV LHC energies. The motivation was to study possible distinctive features of jets and jet quenching in deformed collision systems. HYDJET++ model being modified, provides us an opportunity to study heavy-ion collisions in various geometrical configurations. The collisions in various geometrical configurations are conscious to the initial conditions. Without being imperative on precision, we have tried to make qualitative as well as quantitative study of jets through its observables. Our study involves body-type and tip-type type of collision geometry. We have compared our Monte Carlo HYDJET++ model results with ALICE experimental data \cite{ALICE:2018hza} and AMPT model in string-melting version \cite{kundu2019study} where possible. 

We have presented results for transverse momentum ($p_{T}$) spectra with kinematic range $|\eta|<0.8$ for (0-60)\% centrality of collision dividing them into seven classes such that pseudorapidity distribution of Xe-Xe collisions at 5.44 TeV from HYDJET++ model match ALICE experimental data correspondingly. We present comparison of $p_{T}$-spectra of charged hadrons with and without hard parton scattering or jet part. Minimum biased transverse momentum distribution of charged hadrons shows suitable match with ALICE experimental yield from low to high $p_{T}$ region. However, without contribution from hard parton scattering, $p_{T}$-spectra is very small, hence underpredicts while matching experimental data only at very low $p_{T}$. Transverse momentum decreases as we move from most-central to most-peripheral collisions. This indicates that fireball formed upon colliding nuclei, has higher temperature in central collisions compared to that in peripheral collisions. Yield for tip-tip collisions (higher fireball temperature) is higher than body-body collisions (relatively smaller fireball temperature). The difference between body-body and tip-tip geometrical configurations arises only when hard parton scatterings are involved otherwise yield is same for them.

Average transverse momentum $\langle p_{T} \rangle$ for hydro+jet and without jet part show strong dependence on collision centrality. HYDJET++ model shows a suitable match with the experimental measurements within error bars. However, $\langle p_{T} \rangle$ for only jet part is somewhat centrality independent. This behaviour is in well agreement with results from AMPT model in string-melting version. Nuclear modification factor $R_{AA}$ of charged hadrons is calculated as a function of transverse momentum and collision centrality. Minimum bias $R_{AA}$ of charged hadrons shows a suitable match with ALICE experiment up to 1.0 GeV/c, underestimating data above this $p_{T}$. $R_{AA}$ increases almost 2.22 times for body-body collisions while 1.27 times for tip-tip collisions as we move from most-central to most-peripheral class of collisions, showing a peak at around $p_{T}\simeq$1.5 GeV/c in most central collisions and shifting to $\sim$1.0 GeV/c in most peripheral collisions. Also, the modification factor is 1.32 times higher for body-body collisions than tip-tip collisions. 

In order to study suppression independent of the yield from p-p collisions, relative suppression $R_{CP}$ is studied as a function of  $p_{T}$ and centrality. $R_{CP}$-spectra suitably matches with ALICE experimental data within error bars except at very low $p_{T}$, showing a peak at $p_{T}\simeq$2.25 GeV/c. It increases as we move from most-central to most-peripherals collisions. $\langle R_{CP} \rangle$ is positively correlated to collision centrality, matching ALICE experimental data qualitatively. However, overpredicts quantitatively in central collisions, with a pleasant match as we move towards peripheral collisions.

Thus, our study presented will enlighten the contribution of jets and associated suppression (jet quenching) caused in transverse momentum spectra in deformed Xe-Xe collision systems. The effect of geometry of the colliding system on such smaller systems can be very clearly visible from our work. The fireball created upon collision at such LHC energy and its evolution that too in various geometrical configurations divulge the role of soft and most importantly hard parton scatterings in them.

\vspace{-2ex}
\section*{ACKNOWLEDGEMENTS}
\vspace{-2ex}
We sincerely acknowledge financial support from the Institutions of Eminence (IoE) BHU grant number-6031. Saraswati Pandey acknowledges the financial support obtained from UGC under a research fellowship scheme during the work.

\bibliographystyle{apsrev}

\bibliography{xe}

\end{document}